\documentclass[english]{revtex4-1}
\usepackage[T1]{fontenc}
\usepackage[latin9]{inputenc}
\usepackage{xcolor}
\usepackage{float}
\usepackage{textcomp}
\usepackage{amstext}
\usepackage{graphicx}
\PassOptionsToPackage{normalem}{ulem}
\usepackage{ulem}
\usepackage{babel}

\begin{document}
\title{Influence of the Radial Electric Field on the Shearless Transport
Barriers in Tokamaks}
\author{F.A.Marcus$^{1}$, M. Roberto$^{1}$, I. L. Caldas$^{2}$, K. C. Rosalem$^{1}$,
Y. Elskens$^{3}$}
\affiliation{$^{1}$Departamento de Fisica, Instituto Tecnológico da Aeronáutica,
São José dos Campos, SP 1228-900, Brazil.}
\affiliation{$^{2}$Instituto de Física, Universidade de São Paulo, São Paulo,
SP 05315-970, Brazil.}
\affiliation{$^{3}$Aix-Marseille Université, UMR 7345 CNRS, campus Saint-Jérôme,
case 321, av. esc. Normandie-Niemen, FR-13397 Marseille cedex 20}
\date{08/10/2018}
\begin{abstract}
In tokamaks, internal transport barriers, produced by modifications
of the plasma current profile, reduce particle transport and improve
plasma confinement. The triggering of the internal transport barriers
and their dependence on the plasma profiles is a key nonlinear dynamics
problem still under investigation. We consider the onset of shearless
invariant curves inside the plasma which create internal transport
barriers. A non-integrable drift-kinetic model is used to describe
particle transport driven by drift waves and to investigate these
shearless barriers onset in tokamaks. We show that for some currently
observed plasma profiles shearless particle transport barriers can
be triggered by properly modifying the electric field profile and
the influence of non-resonant modes in the barriers onset. In particular,
we show that a broken barrier can be restored by enhancing non-resonant
modes.
\end{abstract}
\maketitle
\tableofcontents{}

\section{Introduction}

The plasma confinement in tokamaks is limited by particle transport
induced by the electrostatic turbulence \citep{Cavedon_NF2017}. For
some discharges, internal transport barriers reduce this transport
and improve the plasma confinement \citep{Horton_Book12}. Experiments 
show that such barriers appear by modifications of the current profile
using external heating and current drive effects \citep{Wolf_PPCF2003}.
In fact, besides the recent progress to understand this dependence
\citep{Cavedon_NF2017}, the triggering of internal transport barriers
and their dependence with the plasma profiles still
remain a central question to be better understood \citep{Wolf_PPCF2003,Connor_NF2004,Garbet_PPCF2004}. 

Much research has been done on the nature of the transport barrier
in high confinement mode discharges, in a number of tokamaks worldwide,
and the influence of radial electric fields on the particle transport
in magnetically confined fusion plasmas is by now well established
\citep{Viezzer_NF2013}. Specifically, measurements of the radial
electric field indicate that the negative shear region of the $E_{r}$
profile plays a key role in turbulence reduction observed in H-mode,
paving the way towards an improved understanding of the pedestal structure
\citep{Viezzer_NF2013}. So, high-accuracy characterization of the
edge radial electric field can be used to validate transport theory
and identify the onset of transport barriers \citep{Viezzer_NF2013,Severo_NF2009}.
In this context, the $\mathbf{E}\times\mathbf{B}$ shear stabilization
effect has been considered to be the origin of transport barriers
identified in tokamaks \citep{Burrell_PoP1997a}.

On the other hand, the onset of shearless invariant curves inside
the plasma could be a factor responsible for the formation of some
internal transport barriers \citep{Constantinescu_PhyScr2005,Caldas_PPCF2012}.
In fact, these curves act as dikes preventing chaotic particle transport
across them, and so are identified as a kind of shearless transport
barrier. The essentials of a system with shearless transport barriers
are exhibited by a simple symplectic two-dimensional mapping called
standard non-twist map \citep{Negrete_PhyD1996}. As shown in Ref.
\citep{Szezech_Chaos2009} for this map, even after the invariant
surfaces have been broken, the remnant islands may present a large
stickiness that reduces the transport. 

Concerning the context of particle transport in tokamaks, the onset
of these shearless barriers has been proposed to explain the reduction
of transport in tokamaks \citep{Horton_Pop1998} and helimaks \citep{Ferro_PLA2018}.
In fact, in \citep{Horton_Pop1998}, for large aspect ratio tokamaks,
a non-integrable drift model has been proposed to interpret the high
particle transport at the plasma edge as being induced by the electrostatic
turbulence, as caused by the $\mathbf{E}\times\mathbf{B}$ chaotic
radial drift motion of particles. Furthermore, this model has been
applied to identify particle barriers in tokamak experiments \citep{Marcus_PoP08,Caldas_PPCF2012}. 

The model introduced in \citep{Horton_Pop1998} is applied to show
that, for some currently observed plasma profiles, shearless particle
transport barriers can be triggered by alterations on the plasma profiles.
These barriers can appear due to modes present in the turbulence,
and the resonant conditions are determined by the combination of the
safety factor, electric radial field component and the plasma toroidal
velocity profiles. These profiles determine, respectively, the magnetic,
radial electric field and plasma toroidal velocity shears, which are
the relevant control parameters to specify the resonant condition.
We show that enhancing non-resonant waves amplitude may restore shearless
barriers while the resonant modes increase the particle chaotic transport.
We also present examples for which the chaotic particle transport
is reduced by the barrier onset due to slightly modifying the plasma
parameters or even increasing the turbulence level. 

In Section \ref{sec:Drift_Model}, we introduce the drift wave transport
model used in the article. In Section \ref{sec:Transport-barriers},
we present the equilibrium profiles and plasma parameters assumed
in this article, and how the transport barriers are formed. In Sections
\ref{sec:Non-Resonant-Mode} and \ref{sec:Influence-Eletric}, we
analyze numerically the influence of the electric field profile and
non-resonant modes on the barrier formation.

\section{Drift Wave Transport model\label{sec:Drift_Model}}

The model is based on equations of motion that describe particle trajectories
following the magnetic field lines and the electric drift \citep{Horton_Pop1998}.
The particles trajectories are described by the guiding-center equation
of motion

\begin{equation}
\frac{d\mathbf{x}}{dt}=v_{\parallel}\frac{\mathbf{B}}{B}+\frac{\mathbf{E}\times\mathbf{B}}{B^{2}}\label{eq:motion}
\end{equation}

\noindent giving the system of equations,

\[
\frac{dr}{dt}=-\frac{1}{rB}\frac{\partial\tilde{\phi}}{\partial\theta}
\]

\begin{equation}
\frac{d\theta}{dt}=\frac{v_{\parallel}}{r}\frac{B_{\theta}}{B}+\frac{1}{rB}\frac{\partial\tilde{\phi}}{\partial r}-\frac{E_{r}}{rB}\label{eq:System_ODE}
\end{equation}

\[
\frac{d\varphi}{dt}=\frac{v_{\parallel}}{R}
\]

\noindent where $\mathbf{x}=(r,\theta,\varphi)$ is written in local
polar coordinates standing $r$ the radial position, $\theta$ and
$\varphi$ the poloidal and toroidal angles, $R$ is the major plasma
radius, $v_{\parallel}$ is the toroidal velocity of the guiding centers
and $E_{r}$ (r) is the radial electric field profile in equilibrium.

We consider an electric field composed of a radial mean part and a
fluctuating part. Many experiments have shown the simultaneous excitation
of a large spectrum of frequencies $n\omega_{0}$, $n=1,2,..N$, so
the radial electric field of the fluctuating part appears as a wave
spectrum given by \citep{Horton_Pop1998}, 

\begin{equation}
\tilde{\phi}(\mathbf{r},t)=\sum\limits _{L,M,n}\phi_{LMn}\cos(M\theta-L\varphi-n\omega_{0}t+\alpha_{n})\label{eq:Fluctuating_potential}
\end{equation}

where $\tilde{\phi}$ is the fluctuating electrostatic potential such
that $\tilde{\mathbf{E}}=-\nabla\tilde{\phi}$. The spatial electrostatic
mode numbers $L$ and $M$ (respectively toroidal and poloidal) are
assumed to be constant and $\alpha_{n}$ are constant phases that
do not affect the resonant conditions introduced later on.

The magnetic configuration is described by the safety factor $q(r)$,
considering that $B\approx B_{\varphi}\gg B_{\theta}$, which corresponds
to a layer of a large aspect ratio tokamak as in TCABR tokamak ($a/R\simeq0.3$),
where $a$ is the plasma radius. Therefore, the safety factor is calculated
as $q(r)=\frac{rB_{\varphi}}{RB_{\theta}}$. 

The differential equations (\ref{eq:System_ODE}) were normalized
by taking $a$, $B_{0}$ and $E_{0}$ as characteristic length scale,
toroidal magnetic field and mean radial electric field at the plasma
edge. To represent the results in Poincaré sections, we define a normalized
action variable $I\equiv(r/a)^{2}$ and angle variable $\psi_{LM}\equiv M\theta-L\varphi$,
reducing the system of Eqs.(\ref{eq:System_ODE}) to the canonical
pair $(I,\psi).$ Thus, for the normalized variables, the equations
of motion are written as 

\begin{eqnarray}
\frac{dI}{dt} & = & 2M\sum\phi_{n}\sin(\psi-n\omega_{0}t\text{+}\alpha_{n})\label{eq:acao}\\
\frac{d\psi}{dt} & = & \frac{v_{\parallel}(I)}{R}\frac{1}{q(I)}[M-Lq(I)]-\frac{M}{\sqrt{I}}E_{r}(I)\label{eq:angulo}
\end{eqnarray}

Without the fluctuating potential, $\phi_{n}=0$, $I$ is a constant
of motion and the system of equations (\ref{eq:acao}) and (\ref{eq:angulo})
is integrable. The perturbation term consists of a sum of resonant
drift waves, so, for a given wave spectrum, the system is quasi-integrable
and its numerical solutions can be analyzed in phase space $(I,\psi)$.
These solutions give the particle trajectories in phase space typical
of quasi-integrable systems: regular, KAM invariants and islands,
and chaotic trajectories \citep{Escande_PhysRep1985}. The main resonances
can be identified by the islands in phase space. We can analytically
predict the position of primary resonances in phase space by examining
Eq. (\ref{eq:acao}), namely, the resonance location gives the action
$I$ where the wave modes are resonant. The resonances locations are
determined by the action profiles of $v_{\parallel}(I)$, $q(I)$
and $E_{r}(I)$ and by the wave numbers $M$, $L$. 

The islands in the Poincaré maps can be explained by taking the resonance
conditions, which state to the time invariance of the action variable
in Eq. (\ref{eq:angulo}), viz. $\frac{d}{dt}(\psi-n\omega_{0}t)=0$.
Then, the resonance condition is obtained when $(d\psi/dt)/\omega_{0}$
assumes values of the time mode $n$ in Eq. (\ref{eq:angulo}), which
determines the resonant action $I_{n}$: $n=\frac{1}{\omega_{0}}\frac{d\psi}{dt}$.
Taking $\frac{d\psi}{dt}=n\omega_{0}$ and inserting into Eq. (\ref{eq:angulo})
yields the value of I in which the frequency $n\omega_{0}$ is resonant.
So,

\begin{eqnarray}
n\omega_{0} & = & \frac{v_{\parallel}(r)}{R}\frac{1}{q(I)}[M-Lq(I)]-\frac{M}{\sqrt{I}}E_{r}(I)\label{eq:modo_res}
\end{eqnarray}

In the next sections, particle trajectories are obtained by Bulirsch-Stoer
numerical scheme \citep{NR_C_2007} and their intersections in Poincar\'{é}
sections are shown in $(I,\psi)$ planes. We obtain a Poincaré map
by integrating Eqs (\ref{eq:acao}) and (\ref{eq:angulo}) for various
initial conditions. The intersections of the integrated trajectories
are selected at the toroidal section corresponding to instants $t_{j}=j\,2\pi/\omega_{0}$
$(j=0,1,2,...)$. In Poincaré maps, the (nominal) minor plasma radius
lies at $I=1.0$ , but we choose $I$ up to $1.4$ in order to investigate
the particle transport to the chamber wall. 

\section{Shearless Transport barriers\label{sec:Transport-barriers}}

In general, a shearless transport barrier in a two-dimensional dynamical
system is an invariant curve inside a set of invariant closed curves
characterized by a non-monotonical canonical frequency profile. The
shearless barrier corresponds to a quasi-periodic trajectory with
a local extremum frequency \citep{Farazmand_PhyD2014}. Numerical
studies show that the main feature of the shearless barrier compared
with other KAM tori is that such barriers are more robust under time-periodic
perturbations \citep{Negrete_PhyD1996,Szezech_PRE2012}. This kind
of barrier appears in the model considered in this work and has a
dependence on the plasma profiles. Shearless barriers have been well
described in the canonical Hamiltonian systems \citep{Negrete_PhyD1996,Del-Castillo-Negrete_PoP2000,Morrison_PoP2000},
adopted to present this barrier in the chaotic particle transport
in tokamaks. 

In our model, for null perturbing amplitude waves, $\phi_{n}=0$,
the system is integrable, each trajectory is periodic or quasi-periodic
and stays in an invariant line with the initial action $I_{0}$ constant.
In this case, for each iterate in the Poincaré map, the associated
helical angle $\psi$ increases by a constant $\Omega_{0}=\Delta\psi$,
defined as the rotation number, which characterizes the invariant
line. 

In general, for non-vanishing $\phi_{n}$, we have a mixed system
with chaotic trajectories and regular trajectories in invariant lines.
In that sense, the rotation number profile can be an indicator of
the behavior of the trajectories in any region of the phase space.
For the non-integrable case, we can still define a rotation number
for the remaining invariant lines, considering an initial condition
$\psi_{0}$, as the limit $\Omega=\lim_{i\to\infty}(\psi_{i}-\psi_{0})/i$,
where $\psi_{i}$ refers to the $i$-th section. 

To determine the rotation number profile of the remaining invariant
lines, we calculate the invariant rotation number, i.e., the limit
$\Omega$, for initial conditions with a fixed angle variable $\psi_{0}$
and a sequence of action variables $I$. If this profile shows an
extremum, i.e., $d\Omega/dI\cong0$, the point $(I,\psi_{0})$ is
a point in a shearless invariant. In this case, a shearless invariant
curve appears in the phase space keeping the chaotic trajectories
separated in two unconnected domains. The indicated shearless invariant
curve acts as a barrier separating the particle orbits in the phase
space and reducing the particle transport, thus, this shearless curve
acts as an internal transport barrier. Even if this barrier is broken
by perturbing waves, we expect from other maps analyses that the chaotic
orbits may present a large stickiness around the remaining islands,
which reduces the transport \citep{Szezech_PRE2012}.

\begin{figure}[H]
  \begin{centering}
    \includegraphics[scale=0.75]{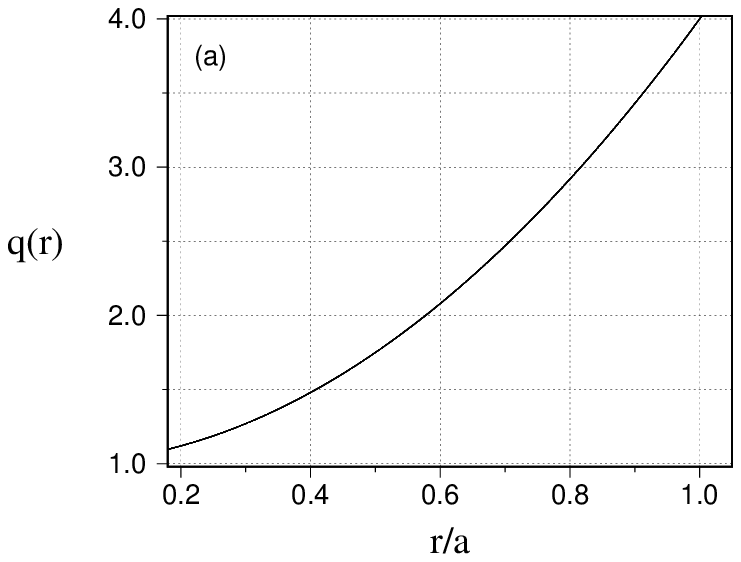}\includegraphics[scale=0.75]{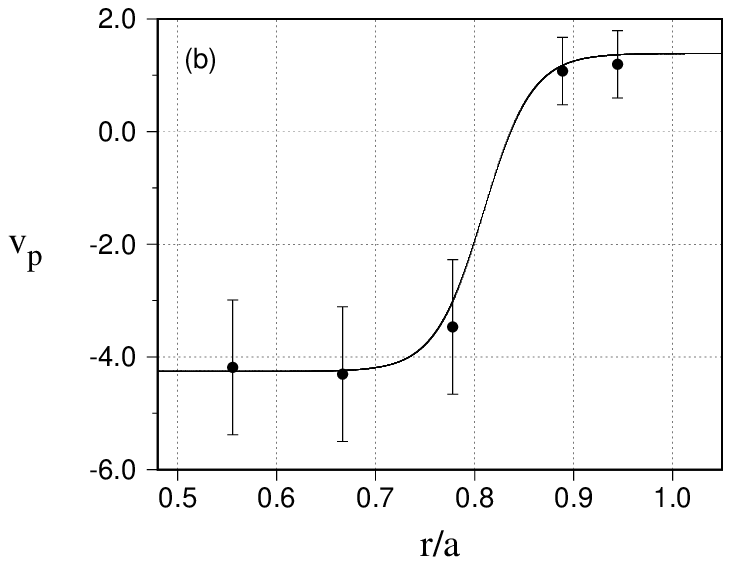}\includegraphics[scale=0.75]{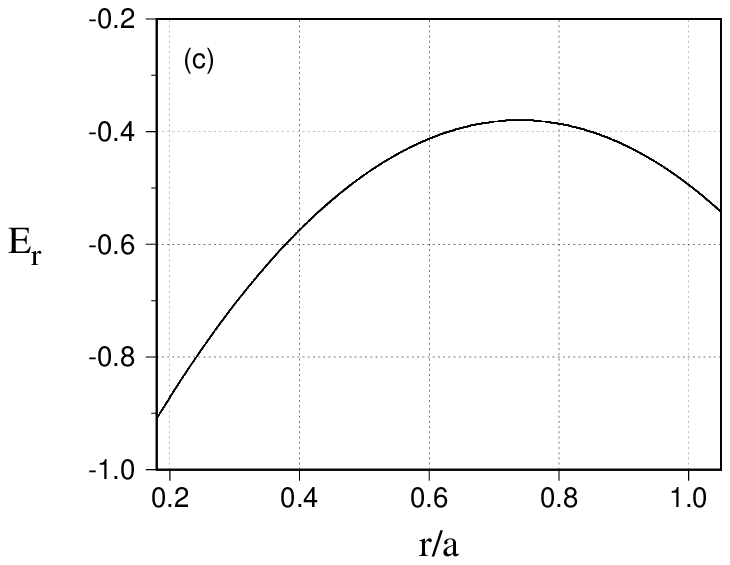}
    \par
  \end{centering}
  \caption{Plasma profiles from tokamak TCABR used in this work. From left to
right, we show a typical profile for safety factor $q(r)$ \citep{Fernandes_Msc2016},
$v_{\parallel}$ with experimental data points \citep{Severo_NF2009}
fitted by a hyperbolic tangent function, and radial electric field
profile in equilibrium \citep{Rosalem2014}. }
  \label{fig:Basic_profiles}
\end{figure}

The existence and location of shearless barriers depend on the $q(I)$,
$v_{\parallel}(I)$ and $E_{r}(I)$ profiles, which are displayed
in Fig.\ref{fig:Basic_profiles}a, \ref{fig:Basic_profiles}b and
\ref{fig:Basic_profiles}c, respectively. These profiles are chosen
similar to those observed in the small tokamak TCABR \citep{Nascimento_NF2005,Severo_NF2009},
but our results can be applied to any tokamak described in a large
aspect ratio approximation. To show how the shear profiles modifications
create transport barriers, numerical simulations are presented for
parameters and profiles taken from the tokamak TCABR. Thus, this paper
presents a conceptual investigation rather than detailed comparisons
with specific experiments performed in any tokamak.

TCABR's safety factor is described by $q(r)=1.0+3.0(r/a)^{2}$, where
$a$ stands for the plasma radius \citep{Fernandes_Msc2016}. We choose
to describe the parallel velocity profile as $v_{\parallel}(r)=-1.43+2.82\tanh(20.3r/a-16.42)$,
which is a fit chosen from experimental data points, as displayed
in Fig. \ref{fig:Basic_profiles}b. The equilibrium radial field $E_{r}$
was chosen to be non-monotonic according to $E_{r}(r)=3\alpha(r/a)^{2}+2\beta(r/a)+\gamma$,
with $\alpha=-0.563$, $\beta=1.250$ and $\gamma=-1.304$, and we
select from the spectrum analysis an frequency
around $10$kHz, which gives us $\omega_{0}=2.673$. The perturbing
electric potential amplitudes $\phi_{n}$ are normalized by $a\,E_{0}$.
 
\begin{figure}[H]
  \centering{}\includegraphics[scale=0.75]{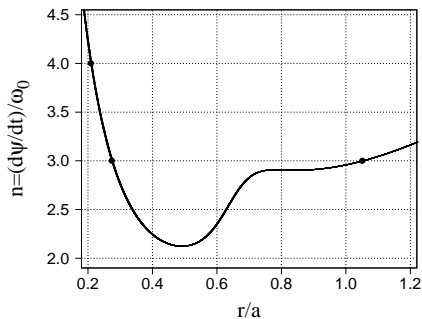}
  \caption{Characteristic curve for resonant modes calculated from Eq.(\ref{eq:modo_res})
using profiles from Fig. (\ref{fig:Basic_profiles}). For integer
$n$, the intersection of the curve with the horizontal lines gives
the position of the island center, which characterizes it as a resonant
mode. We see that the mode $n=3$ has two $I$ values satisfying the
resonance condition, resonant mode $n=4$ appears for only one $I$
value, while $n=2$ is not a resonant mode.}
  \label{fig:resonant_curve}
\end{figure}

The result for the profiles described in Fig. \ref{fig:Basic_profiles}
and spatial wave numbers $M=16$, $L=4$, chosen as typical numbers
in the tokamak wave spectrum at plasma edge \citep{Horton_Pop1998},
into Eq. (\ref{eq:modo_res}) is the resonance profile represented
in Fig. \ref{fig:resonant_curve}. Each point of this curve with an
integer ordinate identifies a mode $n$ which is resonant, i.e., which
generates islands in the Poincaré section. Not only
can we get the mode number, but also the number of centers for each
mode and the radial position ($a\sqrt{I}$) of the centers. As seen
in Fig. \ref{fig:resonant_curve}, we see that the mode $n=3$ has
two islands with centers at $I=(0.27,1.05)$, while $n=4$ has one
center at $I=0.21$. In this way, our study was directed to the interaction
of a doublet of same-frequency resonance modes ($n=3$), a single
resonance mode ($n=4$) and a non-resonant mode ($n=2$).

\begin{figure}[h]
\centering{}\includegraphics[scale=0.75]{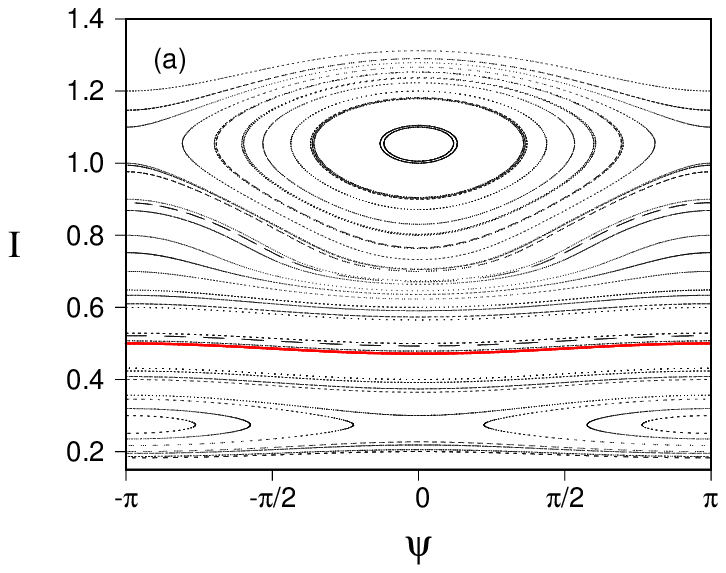}\includegraphics[scale=0.75]{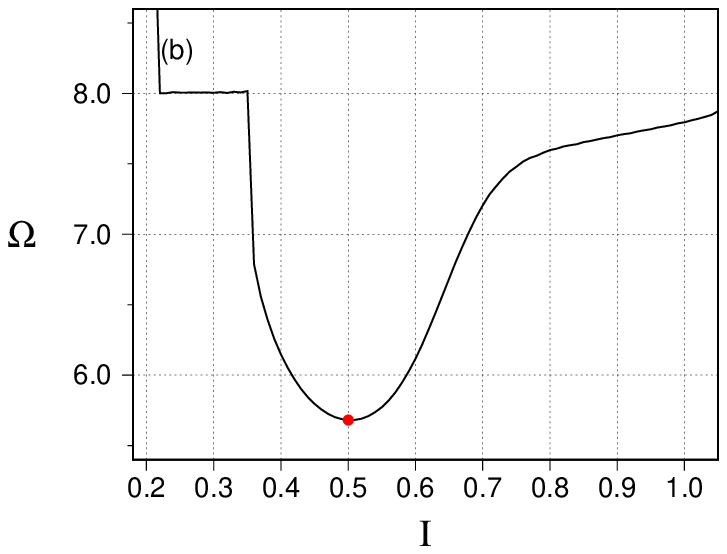}\caption{a) Poincaré map for a single resonant mode $n=3$ with $\phi_{3}=1.0\times10^{-3}$
(normalized by $a\,E_{0}$). The shearless barrier is shown as a red
line. In (b) we calculate the rotation number, for which the local
minimum stands for the shearless barrier position, which is at $I=0.49$.}
\label{fig:Poincare_phi3}
\end{figure}

To have a clear image on how the chosen modes are superimposed and
whether each of them possesses a shearless barrier, the first approach
is to see their aspect individually on a Poincaré section and determine
the rotation number profile. To see the aspect of perturbing period-two
resonant mode $n=3$ and its barrier position, we display the numerical
solution in Fig. \ref{fig:Poincare_phi3}a, with the shearless barrier
highlighted by a red line (color on-line). Due to the chosen equilibrium
profiles, the resonant mode creates islands in two different ranges
in phase space determined by the resonance conditions, as presented
in Fig. \ref{fig:resonant_curve}. The rotation number profile $\Omega(I)$
was calculated with initial angle at $\psi_{0}=-\pi$, as shown in
Fig. \ref{fig:Poincare_phi3}b, and the barrier position is indicated
by a red dot (color on-line), the local minimum. 

\begin{figure}[H]
\centering{}\includegraphics[scale=0.75]{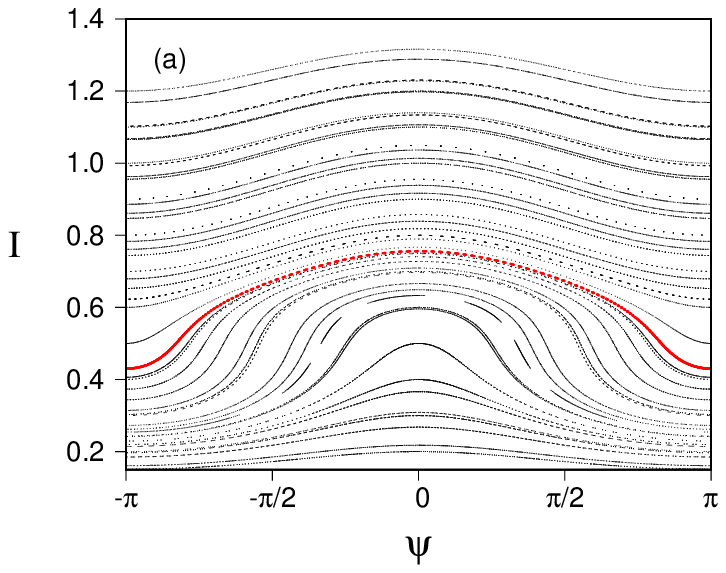}\includegraphics[scale=0.75]{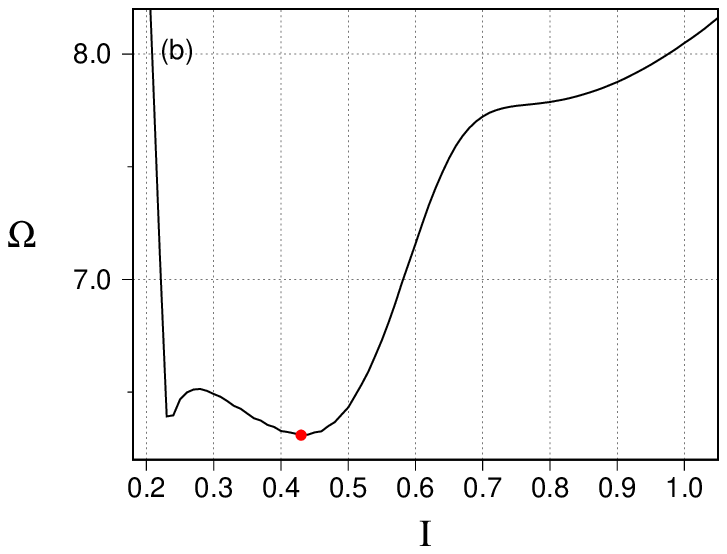}\caption{Poincaré portrait for a non-resonant mode $n=2$ with $\phi_{2}=3.6\times10^{-3}$
(the perturbing potential $\phi_{2}$ is normalized by $a\,E_{0}$).
The important aspect is to show that, if the mode is not resonant,
the contribution of this mode is for barrier formation, which decreases
the chaos when more resonant modes are acting on the system. }
\label{fig:Poincare_phi2}
\end{figure}

It is known that no islands are present if a mode is non resonant.
From a set of invariant lines with the initial action $I_{0}$ constant,
it is important here to see how ``wavy'' these invariant lines become
in the presence of single frequency, if there is a shearless barrier
and where its position on phase space is. The invariant lines of mode
$n=2$ are depicted on Poincaré section followed by its rotation number
in Fig. \ref{fig:Poincare_phi2}. As before, we detected a local extremum
in rotation number, Fig. \ref{fig:Poincare_phi2}b, characterizing
a shearless barrier at $I=0.43$ for $\psi_{0}=-\pi$.

Having presented the configuration on the Poincaré portrait of a resonant
mode of period two and a non resonant and the presence of a shearless
barrier on each of them, the next step is to verify the outcome from
the non linear interactions between three modes on chaos and on transport
barriers formation.

\section{Non Resonant Mode Amplitude\label{sec:Non-Resonant-Mode}}

\begin{figure}[H]
\begin{centering}
\includegraphics[scale=0.7]{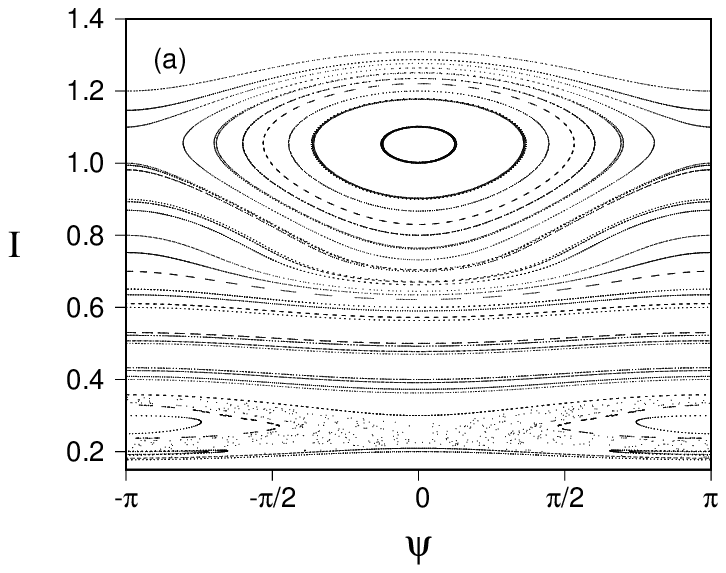}\includegraphics[scale=0.7]{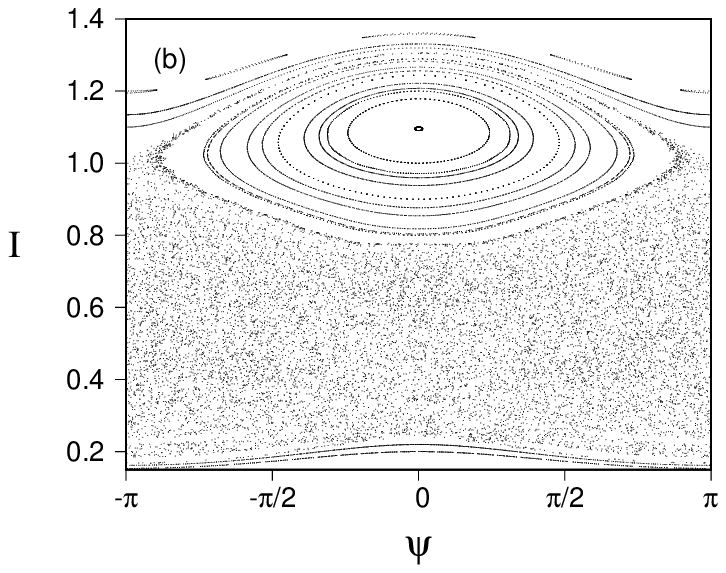}\includegraphics[scale=0.7]{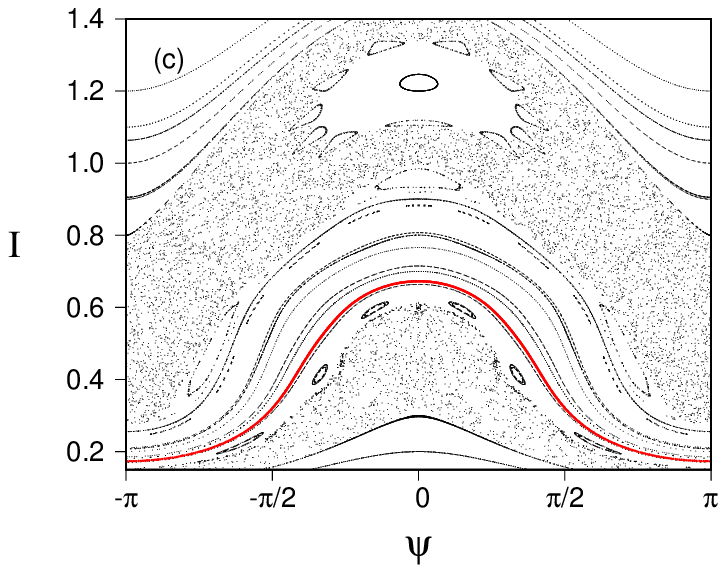}
\par\end{centering}
\caption{Poincaré sections with three modes $n=(2,3,4)$. In all panels $\phi_{3}=1.0\times10^{-3}$
and $\phi_{4}=0.12\times10^{-3}$ . On the left panel $\phi_{2}=0$,
where the chaotic region is delimited by the islands overlapping at
$I\simeq0.25$. For $\phi_{2}=3.6\times10^{-3}$, the middle panel
exhibits chaos at $0.2<I<1.0$. Raising the value of the electric
potential for the non-resonant mode $n=2$ to $\phi_{2}=18\times10^{-3}$,
the shearless barrier split the chaotic area (on the right panel).}
\label{fig:Making_phi2_bigger}
\end{figure}

This section discusses the role of the non-resonant mode on chaotic
mappings obtained from Eqs (\ref{eq:acao}) and (\ref{eq:angulo})
integrated with three modes. With only one resonant mode of period
two, we have two islands separated by invariant curves with a shearless
barrier, like $n=3$ displayed in Fig. \ref{fig:Poincare_phi3}. Increasing
the amplitudes in this kind of system with only one resonant mode
generates no visible chaotic region. For chaos to occur, there must
be an overlap between two islands of different modes, as in the case
of $n=3$ and $n=4$ where the centers are near, and this effect can
be seen on Fig. \ref{fig:Making_phi2_bigger}a. 

The role of the non resonant perturbation $n=2$ is
illustrated in Fig. \ref{fig:Making_phi2_bigger}b and c. Figure \ref{fig:Making_phi2_bigger}b
displays a Poincaré section with a combination of three modes, $n=(2,3,4)$,
with amplitudes given by $\phi_{n}=(3.6,1.2,0.12)\times10^{-3}$,
amplitudes that correspond to those obtained in spectral analysis
on a typical tokamak discharge \citep{Nascimento_NF2005,Nascimento_NF2007}.
The chaos on the Poincaré section results mainly from the overlapping
of the modes $n=3$ and $n=4$. So the non-resonant mode $n=2$ is
contributing to spread the chaos over a larger region beyond the overlapping
region, which means that the shearless barrier is broken.

Increasing the non-resonant mode amplitude to $\phi_{2}=18\times10^{-3}$,
the result is a Poincaré map with the chaotic region splitted by a
shearless barrier, as seen on Fig. \ref{fig:Making_phi2_bigger}c.
Making the non-resonant mode a dominant mode, its contribution is
to establish the transport barrier and reduce the chaotic area. We
want to point out that we can recover the shearless barrier setting
$\phi_{2}\sim4.0\times10^{-3}$; the large value of $\phi_{2}$ was
chosen to make clear the structure brought up by the non resonant
mode.

In summary, a non-resonant mode can be responsible for broadening
the chaotic area and also for the barrier formation. To conclude so,
we consider, for the equilibrium profiles of Fig. \ref{fig:Basic_profiles},
a combination of two resonant modes $n=3$ and $n=4$ and a non resonant
mode $n=2$, as shown in Fig. \ref{fig:resonant_curve}. For $\phi_{2}=3.6\times10^{-3}$,
the Poincaré map of Fig. \ref{fig:Making_phi2_bigger}a shows a resonant
island and a chaotic area. However, increasing the non resonant mode
amplitude for $\phi_{2}=18\times10^{-3}$, we get a Poincaré map with
a shearless barrier. So, the barrier onset is associated to the increasing
of the non resonant mode amplitude. On the other hand, resonant modes
create islands and the islands superposition gives rise to a chaotic
area between the islands. Thus, the influence of $n=2$ on the portrait
is to reduce the chaos, and more important, it also reinstates the
shearless barrier, indicated by the red curve, dividing the chaotic
region in two parts. The higher amplitude perturbation introducing
order can be interpreted as a consequence of non-local perturbation
introduced by the non-resonant mode $n=2$ which alters the global
phase space configuration and induces a bifurcation with a shearless
curve. 

A similar effect has been reported to explain the reversed field pinch
stability induced by a non-resonant perturbation in the magnetic field.
Namely, in the RFX experiment, a non-resonant perturbation reduced
chaos by inducing a bifurcation which modified the phase space configuration,
from a multi-helicity to a single helicity state \citep{Escande_PRL2000,Lorenzini_NatPhys2009}.
Another similarity is found in stellarators, as in the Wendelstein
7-X, for which a carefully tailored topology of nested magnetic surfaces
needed for good plasma confinement is realized even with magnetic
field errors caused by the placement and shapes of the planar coils
\citep{Pedersen2016}.

\section{Influence of the Electric Field Profile\label{sec:Influence-Eletric}}

Since the discovery of the L-H transition in ASDEX \citep{Wagner_PRL1982},
many theoretical and experimental studies have confirmed the importance
of the radial electric field for the formation of internal transport
barriers (ITBs) associated with the $\mathbf{E}\times\mathbf{B}$
velocity shear in magnetic confinement devices \citep{Burrell_PoP1997a,Connor_NF2004}
(see also references therein). Depending on the equilibrium profiles,
small changes on the radial electric field profile may contribute
to the transport barriers onset. However, based on the particle guiding-center
model proposed in \citep{Horton_Pop1998}, we conjecture that transport
barriers may be generated not only due to electric field alterations
but rather whenever appears a local shearless condition, depending
on the $q,v_{\parallel}$ profiles. 

\begin{figure}[H]
\begin{centering}
\includegraphics[scale=0.75]{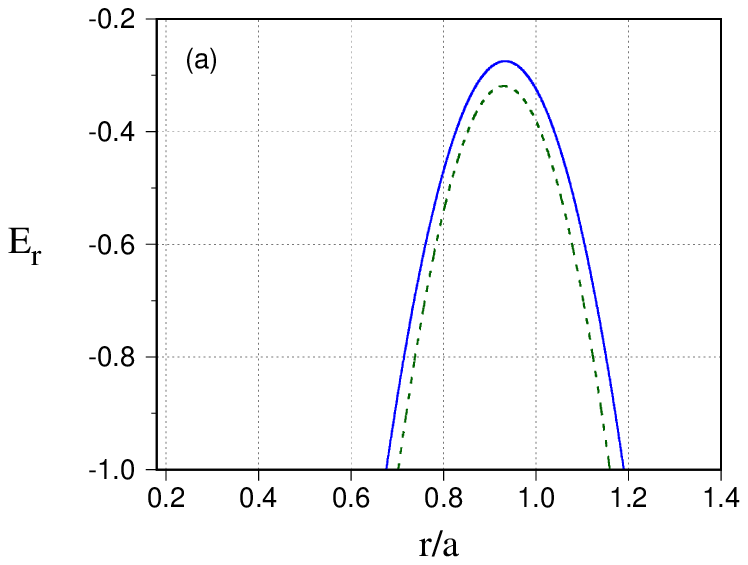}\includegraphics[scale=0.75]{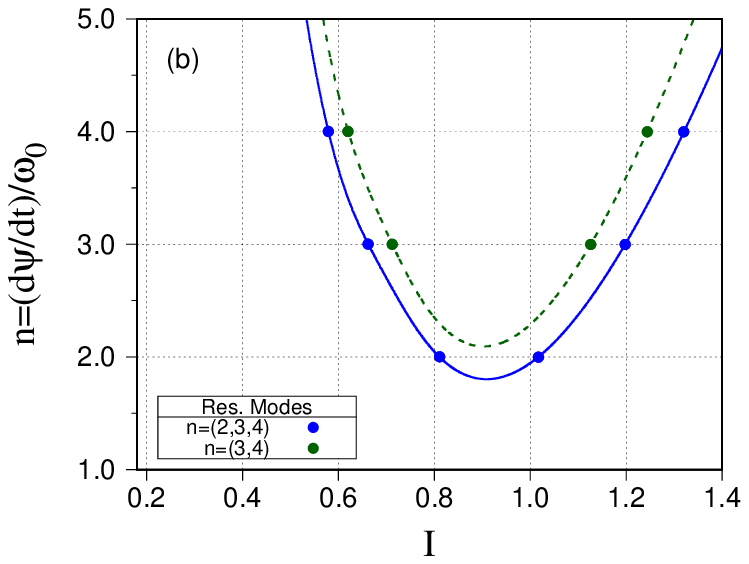}
\par\end{centering}
\caption{a) Two possible profiles for $E(r)$. b) Resonance conditions for
profiles shown in (a). One profile has three resonant modes ($n=2,3,4$),
represented by solid blue line, and the other has only two resonant
modes ($n=3,4$), dashed green line.}
\label{fig:delta_Er_and_resonants_modes} 
\end{figure}

To illustrate our conjecture, we choose two $E_{r}(r)$ profiles presented
in Fig. \ref{fig:delta_Er_and_resonants_modes}a (a different profile
from that used in previous sections), with corresponding resonance
profiles shown in Fig. \ref{fig:delta_Er_and_resonants_modes}b, according
to its corresponding dash pattern. This small change in the radial
electric field is achieved by setting an electrode in which is applied
an electric potential difference, as it has been done in TCABR \citep{Nascimento_NF2005,Nascimento_NF2007}.
The most important alteration is that one profile has three resonant
modes ($n=2,3,4$), represented by solid blue line, and the other
has only two resonant modes ($n=3,4$), dashed green line. 

\begin{figure}[H]
\begin{centering}
\includegraphics[scale=0.75]{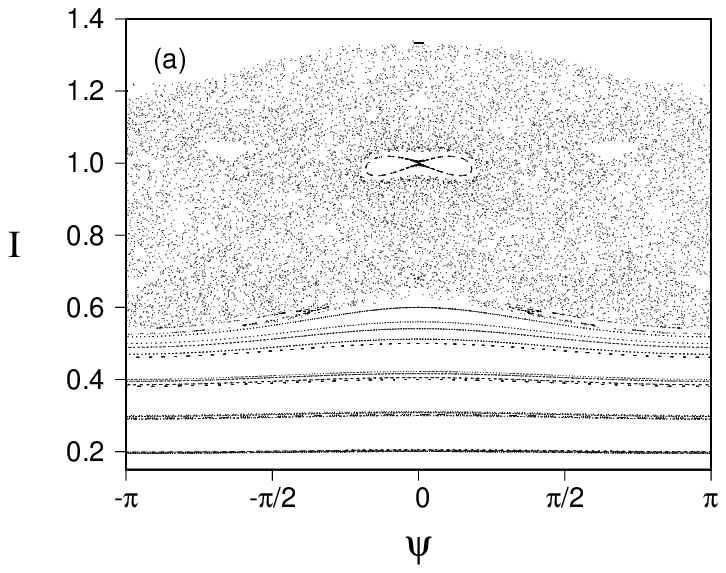}\includegraphics[scale=0.75]{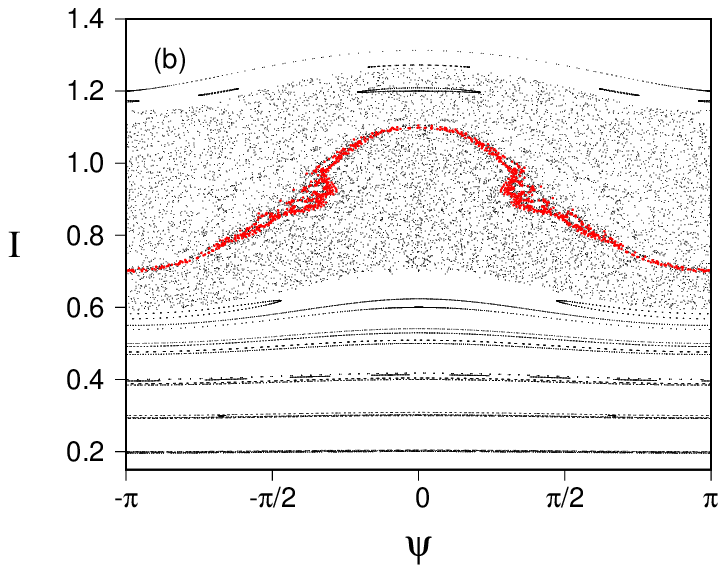}
\par\end{centering}
\caption{Poincaré section for the two electric field profiles given by \ref{sec:Influence-Eletric}
with electric potential $\phi_{n}=(5.9,1.2,0.12)\times10^{-3}$. $n=(2,3,4)$.
In (a) the three modes are resonant, while in (b) only $n=2$ is a
not resonant mode.}

\label{fig:Poincare_modos_234_34}
\end{figure}

In Fig. \ref{fig:Poincare_modos_234_34}, we see how the two resonance
conditions modify the Poincaré section. Figure \ref{fig:Poincare_modos_234_34}a
is obtained for three resonant modes profile, while Fig. \ref{fig:Poincare_modos_234_34}b
is obtained for two modes. In Fig. \ref{fig:Poincare_modos_234_34}b,
we can identify a barrier that is created once $n=2$ is not anymore
a resonant mode. The small change in $E(r)$ is sufficient to suppress
the resonance condition of the $n=2$ mode, opening the possibility
of a shearless bifurcation seen in Fig. \ref{fig:Poincare_modos_234_34}b.
In this example, the shearless barrier is destroyed if the three modes
are resonant, but it is present if the $n=2$ becomes non-resonant
due to the electric field profile modification. Moreover, this small
modification on the $E_{r}(r)$ profile can occur during a plasma
discharge and produce such a bifurcation with the barrier onset. 

In general, the electrostatic turbulence consists of a broadband spectra
with a mixing of resonant and non-resonant modes. We show here that
some of the \ref{fig:Poincare_phi3} base modes may affect the existence
of barriers once they are sensitive to small changes in the electric
field profile. 

\section{Conclusions}

In our investigation, we apply a model, described by a two-dimensional
symplectic drift map proposed to numerically integrate orbits on the
long transport time scales, avoiding long integration times of the
differential equations typically found for the exact guiding-center
orbits in tokamaks. For typical tokamak equilibrium profiles and spectral
potential, we determine the wave resonance conditions. As expected,
the chaotic region and the particle transport in phase space depend
on the resonant wave amplitudes and the equilibrium shear determined
by the magnetic, electric field and velocity profiles. Within this
model we show numerical examples of the shearless barrier onset that
may occur during the tokamak discharges. 

First we show how the increasing of a non-resonant wave amplitude
can create a shearless transport barriers. This occurs because increasing
the non-resonant wave amplitude modifies the phase space and induces
a bifurcation with a shearless curve. 

After that, we investigate the triggering of shearless particle transport
barriers in tokamaks as a consequence of modifications on the plasma
equilibrium profiles compatible with those commonly observed in tokamaks.
Our results indicate that this barrier triggering could be commonly
observed in tokamaks.

We conjecture that the examples of shearless barrier onset could be
observed in some tokamak discharges during which the wave amplitudes
and the equilibrium shear are spontaneously slightly modified. 
\begin{acknowledgments}
The authors thank the financial support from the Brazilian Federal
Agencies CNPq, grants No. 457030/2014-3 and N° 446905/2014-3, PNPD
CAPES Program, and the São Paulo Research Foundation (FAPESP, Brazil)
under grants No. 2011/19296-1 and N° 2015/16471-8. YE enjoyed the
hospitality of the plasma physics and oscillations control group of
the University of São Paulo and support from CAPES-COFECUB grant Ph
908/18. ILC thanks the hospitality during his stay at Aix-Marseille
Université. We thank Prof. D. Escande for his suggestions.
\end{acknowledgments}

\bibliographystyle{aipnum4-1}

\end{document}